\documentclass[
aps,pra,
twocolumn,
showpacs,
superscriptaddress,
floatfix,
]{revtex4-1}
\usepackage[dvipdfmx]{graphicx}
\usepackage{dcolumn}
\usepackage{bm}
\usepackage{ulem}
\usepackage{amsmath}
\usepackage{amssymb}
\usepackage{txfonts}
\usepackage{hyperref}
\usepackage{color} 
\hypersetup{
  colorlinks=true,
  linkcolor=[rgb]{0.85,0.00,0.15},
  citecolor=[rgb]{0.0,0.15,0.95},
  urlcolor=[rgb]{0.0,0.15,0.95},
  setpagesize=false
}

\begin{document}

\title{Novel MAB phases and insights into their exfoliation into 2D MBenes}

\author{Mohammad Khazaei
\footnote{E-mail address: khazaei@riken.jp}}
\affiliation{Computational Materials Science Research Team, 
RIKEN Center for Computational Science (R-CCS), Kobe, Hyogo 650-0047, Japan}

\author{Junjie Wang}
\affiliation{State Key Laboratory of Solidification Processing, Northwestern
Polytechnical University, Xi'an, Shaanxi 710072, People's Republic of China}
\affiliation{International Center for Materials Discovery, School of Materials
Science and Engineering, Northwestern Polytechnical University, Xi'an, Shaanxi 710072, People's Republic of China}

\author{Mehdi Estili}
\affiliation{Research Center for Functional Materials (RCFM), National Institute for Materials Science (NIMS), Tsukuba 305-0047, Japan}

\author{Ahmad Ranjbar\footnote{Current address: Dynamics of Condensed Matter and Center for 
Sustainable Systems Design, Chair of Theoretical Chemistry, University of Paderborn, Warburger Str. 100, 
D-33098 Paderborn, Germany}}

\affiliation{Computational Materials Science Research Team, 
RIKEN Center for Computational Science (R-CCS), Kobe, Hyogo 650-0047, Japan}

\author{Shigeru Suehara}
\affiliation{International Center for Materials Nanoarchitectonics, National Institute for Materials Science (NIMS), 1-1 Namiki, Tsukuba 305-0044, Ibaraki, Japan}
\author{Masao Arai}
\affiliation{International Center for Materials Nanoarchitectonics, National Institute for Materials Science (NIMS), 1-1 Namiki, Tsukuba 305-0044, Ibaraki, Japan}

\author{Keivan Esfarjani}
\affiliation{Departments of Mechanical and Aerospace Engineering, Physics, and Materials Science and Engineering,University of Virginia, 122 Engineer's Way, Charlottesville,VA 22904, USA}

\author{Seiji Yunoki}
\affiliation{Computational Materials Science Research Team, RIKEN Center for Computational Science (R-CCS), Kobe, Hyogo 650-0047, Japan}
\affiliation{Computational Condensed Matter Physics Laboratory, RIKEN Cluster for Pioneering Research (CPR), Wako, Saitama 351-0198, Japan}
\affiliation{Computational Quantum Matter Research Team, RIKEN Center for Emergent Matter Science (CEMS), Wako, Saitama 351-0198, Japan}

\date{\today}

\begin{abstract}
Considering the recent breakthroughs in the synthesis of novel two-dimensional (2D) materials from layered bulk structures, 
ternary layered transition metal borides, known as MAB phases, have come under scrutiny as a means of obtaining novel 2D transition 
metal borides, so-called MBene.
Here, based on a set of phonon calculations, we show the dynamic stability of many Al-containing MAB phases, 
MAlB (M = Ti, Hf, V, Nb, Ta, Cr, Mo, W, Mn, Tc), M$_2$AlB$_2$  (Sc, Ti, Zr, Hf, V, Cr, Mo, W, Mn, Tc, Fe, Rh, Ni),
M$_3$Al$_2$B$_2$ (M = Sc, T, Zr, Hf, Cr, Mn, Tc, Fe, Ru, Ni), 
M$_3$AlB$_4$ (M = Sc, Ti, Zr, Hf, V, Nb, Ta, Cr, Mo, W, Mn, Fe), 
and M$_4$AlB$_6$ (M = Sc, Ti, Zr, Hf, V, Nb, Ta, Cr, Mo). 
By comparing the formation energies of  
these MAB phases with those of their available competing binary M$-$B and M$-$Al, and ternary M$-$Al$-$B phases, 
we find that some of the Sc-, Ti-, V-, Cr-, Mo-, W-, Mn-, Tc-, and Fe-based MAB phases 
could be favorably synthesized in an appropriate experimental condition.
In addition, by examining the strengths of various bonds in MAB phases via crystal orbital Hamilton population 
and spring constant calculations, 
we find that the B$-$B and then M$-$B bonds are stiffer than the M$-$Al and Al$-$B bonds.
The different strength between these bonds implies the etching possibility of Al atoms from MAB phases, 
consequently forming various 2D MB, M$_2$B$_3$, and M$_3$B$_4$ MBenes. 
Furthermore, we employ the nudged elastic band method to investigate the possibility of the structural phase 
transformation of the 2D MB MBenes into graphene-like boron sheets sandwiched between transition metals 
and find that the energy barrier of the transformation is less than $0.4$ eV/atom. 
\end{abstract}

 \maketitle
\section*{Introduction}

The possibility of chemical exfoliation of layered bulk structures has brought great hope to synthesis novel 2D materials with unique electronic and mechanical properties in the future. 
 In this regard, many members of the family of layered transition metal carbides and nitrides, known as MAX phases,\cite{M.W.Barsoum2000,M.Khazaei2014_1,M.Ashton2016_1} have been
 exfoliated into 2D transition metal carbides.~\cite{M.Naguib2011,M.Naguib2012} 
 MAX phases have a general chemical formula of M$_n$AX$_{n+1}$, where 
 M is a transition metal (Sc, Ti, Zr, Hf, V, Nb, Ta, Cr, and Mo), A is an element from groups IIIA--VIA in 
 the periodic table ({\it i.e.}, Al, Ga, Si, Ge, Sn, Pb, P, As, Bi, S, Se, and Te), and X stands for carbon/nitrogen.\cite{M.W.Barsoum2000,M.Khazaei2014_1,M.Ashton2016_1} Recently, more complex MAX phases with ordered double transition metals 
 have also been developed.\cite{B.Anasori2015_1,R.Meshkian2017,M.Dahlqvist2018}  
 In MAX phases, generally, the strength of M$-$A bonds are relatively weaker than the M$-$X bonds.\cite{M.Khazaei2018_1} 
 Such bond strength characteristics make it possible to break the M$-$A bonds using appropriate 
 chemical solutions while M$-$X bonds remain almost intact.\cite{M.Khazaei2018_1} In other words, by the chemical treatment, the ``A'' atoms are washed out from the MAX phases while the backbones, \textit{i.e.},  
the multilayers of 2D M$_n$X$_{n+1}$, remains stable.  By ultrasonication, the multilayers can be separated into monolayers, named as MXenes.~\cite{M.Naguib2011,M.Naguib2012} 
For examples, Ti$_2$AlC, Ti$_2$AlN, V$_2$AlC, Nb$_2$AlC, Ti$_3$AlC$_2$, Ti$_3$SiC$_2$,  
Ti$_4$AlN$_3$, V$_4$AlC$_3$, Nb$_4$AlC$_3$, Ta$_4$AlC$_3$, Mo$_2$ScAlC$_2$, Mo$_2$TiAlC$_2$, Mo$_2$Ti$_2$AlC$_3$, (Mo$_{2/3}$Sc$_{1/3}$)$_2$AlC, (W$_{2/3}$Y$_{1/3}$)$_2$AlC,  and many others have 
already been exfoliated into their corresponding 2D MXenes.\cite{M.Naguib2011,M.Naguib2012,M.Naguib2013,M.Ghidiu2014,P.Urbankowski2016,B.Soundiraraju2017,M.Alhabeb2018,M.H.Tran2018,B.Anasori2015_1,R.Meshkian2017,M.Dahlqvist2018}
Additionally, the above chemical technique has been successfully applied to exfoliate non-MAX phase structures such as ScAl$_3$C$_3$,
Mo$_2$Ga$_2$C, Zr$_3$Al$_3$C$_5$, and Hf$_3$Al$_4$C$_6$ to obtain 2D ScC$_x$, Mo$_2$C, Zr$_3$C$_2$, and Hf$_3$C$_2$, respectively.\cite{R.Meshkini2015,J.Zhou2016,J.Zhou2017,J.Zhou2019} More recently, MoAlB with double Al layers, a member of transition metal boride phases, known as MAB phases, 
has also been partially etched into 2D MoB.\cite{L.T.Alameda2017,L.T.Alameda2018} Also 2D CrB was obtained from chemical exfoliation of MAB phase Cr$_2$AlB$_2$.\cite{H.Zhang2018,H.Zhang2019} 
In analogous to the MAX phases, where after etching the ``A'' elements their corresponding 2D materials were named as MXenes, the 2D transition metal borides, which are obtained by etching the ``A'' elements from the MAB phases, 
are called MBenes.  
It is noteworthy that the electronic and magnetic applications of 2D transition metal borides are growing very fast.~\cite{J.Wang2017,J.Yu2018,Y.Shao2018,G.Yuan2019,B.Zhang2019}
 Recent calculations indicate that 2D MBenes may have applications for Li- and Na-ion batteries,\cite{Z.Guo2017,T.Bo2019}, 
electrocatalysis,\cite{L.T.Alameda2017,Z.Guo2017} and magnetic~\cite{Z.Jiang2018} devices. 
Similar to 2D MXenes,\cite{M.Khazaei2017, M.Khazaei2019,B.Anasori2017_1,B.M.Jun2018,J.Pang2019,A.L.Ivanovskii2013,N.K.Chaudhari2017,J.Zhu2017,H.Wang2018,X.Li2018,X.Zhang2018,Y.Zhang2018,K.Hantanasirisakul2018,H.Lin2018} it is expected MBenes may find many potential applications in electronic, optoelectronic, and energy devices in the future. 

 \begin{figure*}[t]
\centering
\includegraphics[scale=0.2]{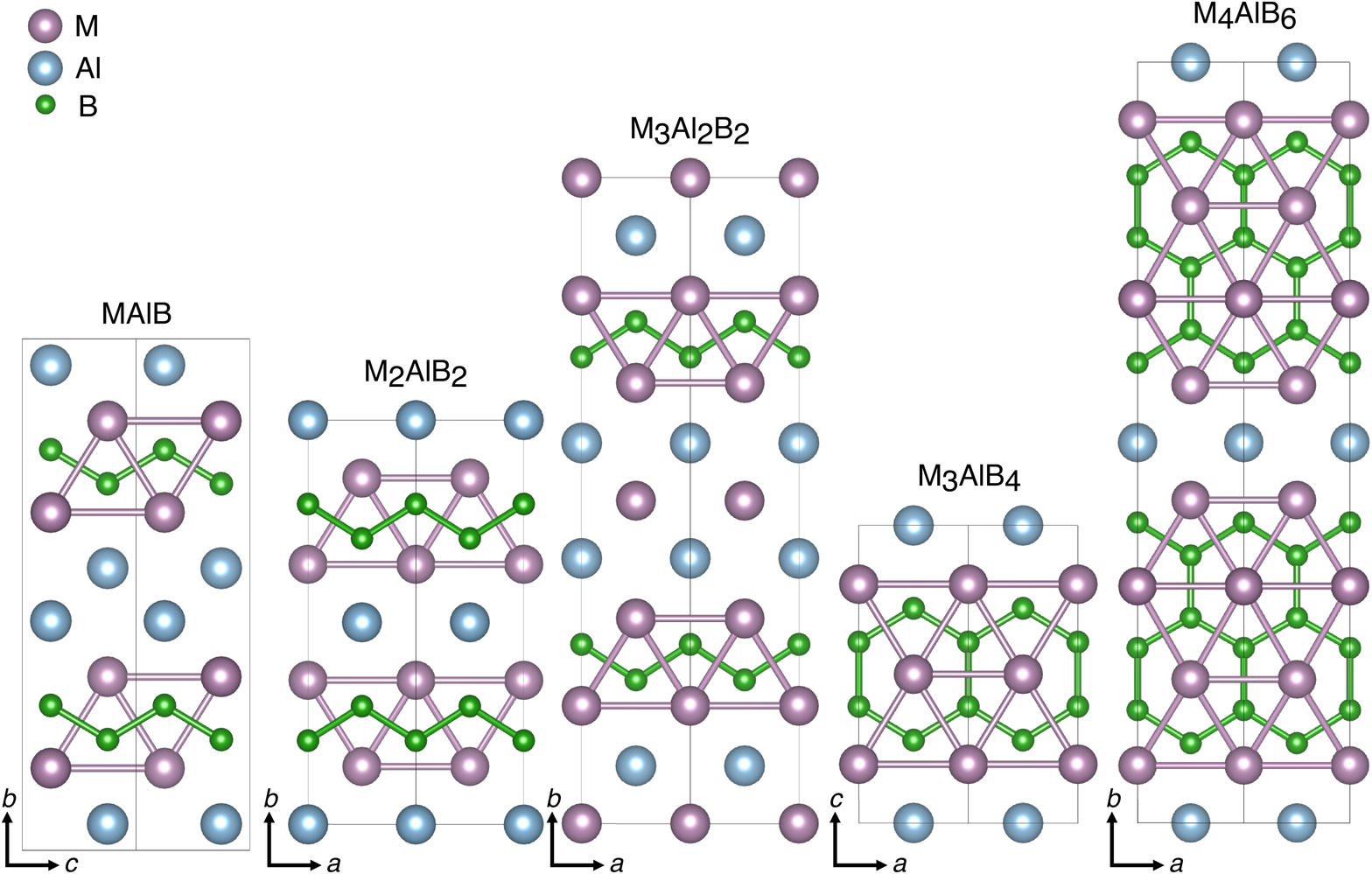}
  \caption{Crystal structures of various MAB phases.  
  M, Al and B atoms are denoted by purple, blue, and green spheres, respectively. 
  $a$, $b$, and $c$ directions are indicated by arrows.}
  \label{fig:mabphase}
\end{figure*}

As listed above, many Al-containing MAX phases have already been exfoliated into 2D MXenes. This motivated 
experimentalists to focus on the Al-containing MAB phases as promising candidates 
for obtaining new 2D transition metal borides.
MAB phases are orthorhombic crystals with the chemical formula of MAlB, M$_2$AlB$_2$, M$_3$Al$_2$B$_2$, 
M$_3$AlB$_4$, and M$_4$AlB$_6$, shown in Figure~\ref{fig:mabphase}. 
Experimentally, MAlB (M = Mo and W), M$_2$AlB$_2$ (M = Cr, Mn, and Fe), Ru$_3$Al$_2$B$_2$, Cr$_3$AlB$_4$, 
and Cr$_4$AlB$_6$ have already been synthesized.\cite{W.Jung1986,X.Tan2013,M.Ade2015,P.Chai2015,S.Kota2016,J.Lu2017,S.Kota2017,H.Zhang2018,H.Zhang2019,T.N.Lamichhane2018}
As seen from Figure~\ref{fig:mabphase}, the skeleton of MAlB, M$_2$AlB$_2$, and M$_3$Al$_2$B$_2$ contains 
isolated zigzag chains of boron atoms, while M$_3$AlB$_4$ and M$_4$AlB$_6$ contain double and triple chains 
of boron atoms, respectively, connected to form flat strips with hexagonal boron ring networks. 
In this paper, we extend the family members of MAB phases by studying the dynamic stability and the formation energy 
of MAlB, M$_2$AlB$_2$, M$_3$Al$_2$B$_2$, M$_3$AlB$_4$, and M$_4$AlB$_6$ (M = Sc, Ti, Zr, Hf, V, Nb, Ta, Cr, 
Mo, W, Mn, Tc, Fe, Ru, Co, Rh, and Ni). In excellent agreement with experiments, our formation energy calculations 
indicate that some of the Cr-, Mo-, W-, Mn-, and Fe-based MAB phases can be formed experimentally. 
Although Sc-, Ti-, and V-based MAB phases are slightly metastable compared to the most stable MAB phases 
described above, they have a high probability of being formed experimentally. Bond strength calculations indicate that 
B$-$B and M$-$B are stiffer than the M$-$Al and Al$-$B bonds, which indicates the possibility of etching the Al atoms 
from MAB phases and thus forming 2D MBenes. Furthermore, using the nudged elastic band method, 
we investigate the possibility of the rearrangement of boron chains in 2D MB MBene into a 2D graphene-like boron sheet 
that is sandwiched between the metal layers. This process results in structural phase transformation of the 2D MB 
MBene with rectangular lattice to a hexagonal lattice. The same transformation process may occur in M$_2$B$_3$ 
and M$_3$B$_4$ MBenes, resulting in the formation of double and triple graphene-like boron sheets sandwiched 
between the metal layers.

\begin{table*}
\caption{List of MAB phases that we found dynamically stable based on phonon calculations. 
\textit{a},  \textit{b},	and \textit{c} are the lattice parameters. $\Delta H$ indicates the instability energy of MAB 
phases over competing phases. 
FC$_M$, FC$_{Al}$, FC$_B$, are total force constants on M, Al, and B atoms, respectively. 
$E_{\rm exf}$ is the static exfoliation energy. 
The MAB phases experimentally synthesized are indicated by asterisks ($\star$). The experimental lattice parameters 
of the MAB phases already synthesized are also listed in parentheses. 
}
\scriptsize
\begin{tabular}{lccccccccc}
\hline
MAB phase & \textit{a} (\AA) &  \textit{b} (\AA)	&	\textit{c} (\AA)	& Most competing phases	 & $\Delta H$ (eV/atom)	&	FC$_M$ (eV/\AA$^2$)	&	FC$_{Al}$ (eV/\AA$^2$) &	FC$_B$ (eV/\AA$^2$)	&	$E_{\rm exf}$ (eV/\AA$^2$) \\
\hline

TiAlB	&	3.285	&	14.661	&	3.055	&	TiAl$_2$, TiB2					&	0.126	&	44.830	&	18.707	&	44.224	&	0.142 \\
HfAlB	&	3.455	&	15.516	&	3.155	&     HfB$_2$, HfAl$_2$					&	0.176	&	46.223	&	15.127	&	40.700	&	0.125 \\
VAlB	&	3.091	&	14.201	&	3.005	&	 V$_3$B$_4$, VAl$_3$, Al					&	0.041	&	48.964	&	26.237	&	45.971	&	0.187 \\
NbAlB	&	3.343	&	14.714	&	3.130	&  NbAl$_3$, Nb$_3$B$_4$, NbB$_2$					&	0.090	&	51.156	&	21.889	&	38.443	&	0.176 \\
TaAlB	&	3.330	&	14.638	&	3.107	&	 Ta$_3$B$_4$, Al, TaAl$_3$					&	0.055	&	53.138	&	24.064	&	40.307	&	0.187 \\
CrAlB	&	3.003	&	13.889	&	2.969	&	Cr$_2$AlB$_2$, Al					&	0.008	&	50.379	&	31.043	&	48.607	&	-0.205 \\
MoAlB$^\star$\cite{M.Ade2015}	&	3.216 (3.199~\cite{M.Ade2015})	&	14.025 (13.922~\cite{M.Ade2015})	&	3.110 (3.094~\cite{M.Ade2015})	&	 MoB, MoB$_2$, Mo$_3$Al$_8$ 					&	-0.114	&	56.367	&	29.542	&	39.139	&	0.225 \\
WAlB$^\star$~\cite{M.Ade2015}	&	3.220 (3.202~\cite{M.Ade2015})	&	13.976 (13.906~\cite{M.Ade2015})	&	3.118 (3.102~\cite{M.Ade2015})	&	WB$_2$, WB, WAl$_5$					&	-0.022	&	57.306	&	31.878	&	39.098	&	0.229 \\
MnAlB	&	2.955	&	13.859	&	2.989	&	 MnAl$_6$, Mn$_2$AlB$_2$, MnB$_4$					&	0.065	&	34.122	&	29.090	&	46.380	&	0.159 \\
TcAlB	&	3.025	&	14.099	&	3.168	&	TcAl$_3$, Tc$_2$Al, TcB$_2$					&	0.015	&	38.686	&	31.604	&	34.388	&	0.203 \\
                &                        &                      &                      &                                                                          &                      &                        &                        &                      &                  \\
                 
Sc$_2$AlB$_2$	&	3.178	&	11.724	&	3.614	&	ScAl, ScB$_2$					&	0.041	&	33.538	&	20.963	&	34.806	&	0.095 \\
Ti$_2$AlB$_2$	 &	3.047	&	11.323	&	3.311	&	Ti$_3$B$_4$, TiB$_2$, TiAl					&	0.026	&	43.162	&	25.248	&	44.433	&	0.149 \\
Zr$_2$AlB$_2$	 &	3.192	&	12.012	&	3.605	&	ZrB$_2$, Zr$_2$Al$_3$, Zr$_4$Al$_3$					&	0.080	&	43.720	&	21.206	&	36.494	&	0.115 \\
Hf$_2$AlB$_2$ &	3.166	&	11.799	&	3.553	&	Hf$_4$Al$_3$, HfB$_2$, HfAl$_2$					&	0.078	&	46.087	&	23.137	&	38.012	&	0.129 \\
V$_2$AlB$_2$	&	3.012	&	11.109	&	3.075	&	VB, V$_3$B$_4$, VAl$_3$					&	0.097	&	47.491	&	25.784	&	45.554	&	0.174 \\
Cr$_2$AlB$_2$$^\star$\cite{M.Ade2015}	&	2.924 (2.937~\cite{M.Ade2015})	&	11.051 (11.051~\cite{M.Ade2015})	&	2.934 (2.967~\cite{M.Ade2015})	&	Cr$_3$AlB$_4$, Cr$_7$Al$_{45}$, CrB 					&	-0.044	&	58.667	&	29.668	&	49.742	&	0.216 \\
Mo$_2$AlB$_2$	&	3.076	&	11.543	&	3.146	&	MoB, MoAlB					&	0.012	&	59.637	&	24.739	&	39.849	&	0.210 \\
W$_2$AlB$_2$ 	&	3.085	&	11.584	&	3.138	&	WB$_2$, WB, WAl$_5$					&	0.039	&	63.566	&	26.312	&	40.231	&	0.215 \\
Mn$_2$AlB$_2$$^\star$\cite{M.Ade2015}	&	2.896 (2.918\cite{M.Ade2015})	&	11.074 (11.038\cite{M.Ade2015})	&	2.831 (2.893\cite{M.Ade2015})	&	Mn$_4$Al$_{11}$, MnB, MnB$_4$					&	-0.060	&	45.586	&	31.649	&	49.062	&	0.193 \\
Tc$_2$AlB$_2$ 	&	3.043	&	11.576	&	3.018	&	TcAl$_3$, Tc$_2$Al, TcB$_2$					&	-0.050	&	48.192	&	28.249	&	38.614	&	0.217 \\
Fe$_2$AlB$_2$$^\star$\cite{M.Ade2015}	&	2.917 (2.922\cite{M.Ade2015})	&	11.024 (10.991\cite{M.Ade2015})	&	2.853 (2.856\cite{M.Ade2015})	&	FeAl$_6$, AlB$_2$, FeB					&	-0.078	&	37.618	&	28.990	&	45.971	&	0.181 \\
Ru$_2$AlB$_2$	&	2.955	&	12.849	&	2.829	&	RuB$_2$, Ru$_4$Al$_3$B$_2$					&	0.107	&	42.628	&	19.776	&	37.506	&	0.205 \\
Rh$_2$AlB$_2$	&	3.103	&	12.253	&	2.860	&	RhAl, RhB, B					&	0.263	&	19.849	&	18.892	&	29.707	&	0.134 \\
Ni$_2$AlB$_2$ 	&	2.980	&	11.051	&	2.850	&	Ni$_{12}$AlB$_8$, B, NiAl					&	0.139	&	19.596	&	25.579	&	38.699	&	0.115 \\
                 &                        &                      &                      &                                                                          &                      &                        &                        &                      &                  \\

Sc$_3$Al$_2$B$_2$	&	3.192	&	18.385	&	3.600	&	ScAl, ScB$_2$					&	0.024	&	33.835	&	20.169	&	34.116	&	0.135 \\
Zr$_3$Al$_2$B$_2$	        &	3.180	&	18.335	&	3.691	&	ZrB$_2$, Zr$_2$Al$_3$, Zr$_4$Al$_3$					&	0.098	&	41.153	&	20.197	&	35.366	&	0.143 \\
Hf$_3$Al$_2$B$_2$ 	&	3.150	&	18.073	&	3.638	&	Hf$_4$Al$_3$, HfB$_2$, HfAl$_2$					&	0.100	&	43.180	&	21.860	&	36.989	&	0.149 \\
Cr$_3$Al$_2$B$_2$ 	&	2.936	&	17.313	&	2.951	&	Cr$_2$AlB$_2$, CrAl$_3$, Cr$_2$Al					&	0.038	&	53.461	&	26.445	&	48.447	&	0.213 \\
Mn$_3$Al$_2$B$_2$ 	&	2.851	&	17.863	&	2.833	&	MnAl, Mn$_2$AlB$_2$					&	0.004	&	49.480	&	26.209	&	50.172	&	0.228 \\
Tc$_3$Al$_2$B$_2$ 	&	3.045	&	17.545	&	3.054	&	TcAl$_3$, Tc$_2$Al, TcB$_2$					&	-0.037	&	48.663	&	26.422	&	37.386	&	0.255 \\
Fe$_3$Al$_2$B$_2$	&	2.913	&	16.536	&	2.875	&	Fe$_2$AlB$_2$, FeAl					&	-0.013	&	35.806	&	30.439	&	45.163	&	0.225 \\
Ru$_3$Al$_2$B$_2$$^\star$\cite{W.Jung1986}	&	3.045 (2.967\cite{W.Jung1986}) 	&	17.824 (17.036\cite{W.Jung1986})	&	2.938 (2.965\cite{W.Jung1986})	&	RuB$_2$, Ru$_4$Al$_3$B$_2$					&	0.070	&	33.390	&	28.486	&	32.482	&	0.277 \\
Ni$_3$Al$_2$B$_2$ 	&	2.962	&	16.926	&	2.840	&	Ni$_{12}$AlB$_8$, B, NiAl					&	0.084	&	20.469	&	25.573	&	38.568	&	0.202 \\
                 &                        &                      &                      &                                                                          &                      &                        &                        &                      &                  \\																	
                
Sc$_3$AlB$_4$	&	3.157	&	3.565	&	8.630	&	 ScAl, ScB$_2$					&	0.035	&	34.006	&	21.866	&	35.629	&	0.100 \\
Ti$_3$AlB$_4$ 	&	3.039	&	3.296	&	8.254	&	Ti$_3$B$_4$, TiB$_2$, TiAl					&	0.040	&	44.608	&	26.571	&	44.955	&	0.152 \\
Zr$_3$AlB$_4$ 	&	3.187	&	3.595	&	8.711	&	ZrB$_2$, Zr$_2$Al$_3$, Zr$_4$Al$_3$					&	0.052	&	45.443	&	22.684	&	36.802	&	0.125 \\
Hf$_3$AlB$_4$ 	&	3.155	&	3.534	&	8.608	&	Hf$_4$Al$_3$, HfB$_2$, HfAl$_2$					&	0.073	&	47.672	&	24.291	&	39.102	&	0.137 \\
V$_3$AlB$_4$  	&	2.976	&	3.059	&	8.191	&	V$_3$B$_4$, VAl$_3$, Al					&	0.046	&	50.634	&	25.465	&	47.017	&	0.195 \\
Nb$_3$AlB$_4$ 	&	3.126	&	3.314	&	8.609	&	NbAl$_3$, Nb$_3$B$_4$, NbB$_2$					&	0.087	&	49.773	&	20.258	&	38.349	&	0.171 \\
Ta$_3$AlB$_4$	&	3.110	&	3.304	&	8.544	&	Ta$_3$B$_4$, Al, TaAl$_3$					&	0.092	&	50.945	&	21.925	&	39.526	&	0.179 \\
Cr$_3$AlB$_4$$^\star$\cite{M.Ade2015}	&	2.939 (2.956\cite{M.Ade2015})	&	2.939 (2.978\cite{M.Ade2015})	&	8.088 (8.054\cite{M.Ade2015})	&	 Cr$_2$AlB$_2$, CrB, CrB$_4$					&	-0.004	&	57.605	&	30.273	&	48.514	&	0.209 \\
Mo$_3$AlB$_4$	&	3.087	&	3.168	&	8.395	&	 MoB, MoB$_2$, MoAlB					&	-0.027	&	58.688	&	25.948	&	38.797	&	0.196 \\
W$_3$AlB$_4$	&	3.093	&	3.177	&	8.381	&	WB$_2$, WB, WAl$_5$					&	0.084	&	60.745	&	28.368	&	38.852	&	0.170 \\
Mn$_3$AlB$_4$	&	2.937	&	2.835	&	8.153	&	MnB, Mn$_2$AlB$_2$, MnB$_4$					&	0.036	&	45.581	&	30.902	&	46.191	&	0.190 \\
Fe$_3$AlB$_4$	&	2.960	&	2.819	&	8.144	&    Fe$_2$AlB$_2$, FeB, B					&	0.063	&	33.713	&	28.099	&	42.178	&	0.163 \\
                  &                        &                      &                      &                                                                          &                      &                        &                        &                      &                  \\		               				
																						
Sc$_4$AlB$_6$	&	3.154	&	22.721	&	3.551	&     ScAl, ScB$_2$, Sc$_2$Al					&	0.028	&	34.172	&	21.600	&	35.482	&	0.099 \\
Ti$_4$AlB$_6$	       &	3.035	&	21.840	&	3.267	&	Ti$_3$B$_4$, TiB$_2$, TiAl					&	0.036	&	44.752	&	26.494	&	44.963	&	0.156 \\
Zr$_4$AlB$_6$ 	&	3.182	&	22.999	&	3.575	&	ZrB$_2$, Zr$_2$Al$_3$, Zr$_4$Al$_3$					&	0.059	&	45.655	&	22.318	&	36.810	&	0.128 \\
Hf$_4$AlB$_6$ 	&	3.150	&	22.731	&	3.511	&	Hf$_4$Al$_3$, HfB$_2$, HfAl$_2$					&	0.061	&	47.867	&	23.789	&	39.083	&	0.141 \\
V$_4$AlB$_6$  	&	2.973	&	21.617	&	3.050	&	V$_2$B$_3$, Al					&	0.026	&	53.266	&	25.311	&	47.074	&	0.198 \\
Nb$_4$AlB$_6$  	&	3.118	&	22.604	&	3.328	&	NbAl$_3$, Nb$_3$B$_4$, NbB$_2$					&	0.057	&	53.062	&	20.864	&	38.425	&	0.168 \\
Ta$_4$AlB$_6$ 	&	3.099	&	22.511	&	3.312	&	TaB$_2$, Ta$_3$B$_4$, Al					&	0.049	&	55.537	&	22.365	&	39.789	&	0.178 \\
Cr$_4$AlB$_6$$^\star$\cite{M.Ade2015}	&	2.947 (2.952\cite{M.Ade2015})	&	21.325 (21.280\cite{M.Ade2015})	&	2.944 (3.013\cite{M.Ade2015})	&	CrB$_4$, Cr$_3$AlB$_4$, CrB					&	0.008	&	57.827	&	30.251	&	48.589	&	0.200 \\
Mo$_4$AlB$_6$	&	3.080	&	22.091	&	3.195	&	MoB, MoB$_2$, MoAlB					&	-0.021	&	58.159	&	25.808	&	38.828	&	0.181 \\
																				
\hline
\end{tabular}
\label{tab:data}
\end{table*}

\section{Methods of calculations}

First-principles calculations based on density functional theory (DFT) are performed to optimize the atomic 
structures and examine the electronic structures of the MAB phases. 
All calculations are carried out using the Vienna ab initio simulation package (VASP) code.\cite{vasp1996} 
The generalized gradient 
approximation (GGA) using the Perdew-Burke-Ernzerhof (PBE) 
functional is used to compute the exchange-correlation 
energy.\cite{pbe1996} 
The projected augmented wave approach with a plane wave cutoff 
energy of 520 eV is used to construct the wave functions. The atomic positions and lattice constants are 
fully optimized using the conjugate gradient method without imposing any symmetry. 
After the optimization of atomic structures, the maximum 
residual force on each atom is less than 0.001 eV/\AA. In the electronic self-consistency procedure, 
the total energies are converged within 10$^{-7}$ eV/cell.  
For the optimization of MAlB, M$_2$AlB$_2$, M$_3$Al$_2$B$_2$, and M$_4$AlB$_6$ (M$_3$AlB$_4$), 
18$\times$9$\times$18 
(18$\times$18$\times$9)
Monkhorst-Pack {\bf k} points are used.\cite{monkhorst1976} 
Most of the MAB phases are nonmagnetic, except for 
Cr-, Mn-, and Fe-based MAB phases. 
The total energies and properties of the magnetic MAB phases and all 2D MBenes are evaluated using 
spin-polarized calculations. 

The phonon calculations are carried out 
using the PHONOPY package~\cite{phonopy2008} 
along with the VASP.\cite{vasp1996} 
To evaluate the strength of the bonds, it is assumed the atoms in a bond are connected by 
springs.\cite{M.Khazaei2018_1} The spring or force constant is the second derivative of the total energy 
with respect to finite 
displacements of atoms $i$ and $j$ along the $x$, $y$, and $z$ directions, and is described as a 3$\times$3 matrix, 
which is one of the output results of the phonon calculations. 
Since the trace of the force constant matrix is independent of the coordinate system, \textit{i.e.}, 
invariant under a coordinate rotation,\cite{A.J.E.Foreman1957,Y.Liu2014}
we consider the value of the trace of the force constant matrix 
and refer to this scalar quantity as the force constant $F_{ij}$ between atoms $i$ and $j$.\cite{M.Khazaei2018_1,Y.Liu2014} 


As another method for evaluating a bond strength, we also employ the crystal orbital Hamilton population (COHP) 
analysis.\cite{R.Dronskowski1993} 
The COHP analysis is a technique for partitioning 
the band-structure energy into bonding, nonbonding, 
and antibonding contributions using localized atomic basis sets.\cite{R.Dronskowski1993} 
It describes as a hopping-weighted density-of-states between a pair of adjacent 
atoms defined as 
\begin{equation}
\small
{\rm COHP}_{\mu\vec{T},\nu\vec{T'}}(E)=H_{\mu\vec{T},\nu\vec{T'}}\sum_{j,\vec{\bf k}}f_j(\vec{\bf k})C^\ast_{\mu\vec{T},j}  (\vec{\bf k}) C_{\nu\vec{T'},j} (\vec{\bf k}) \delta(\epsilon_j(\vec{\bf k})-E),
\label{eq:COHP}
\end{equation} 
where $H_{\mu\vec{T},\nu\vec{T'}}$ is the matrix element of the Hamiltonian matrix $H$ and 
$C_{\mu\vec{T},j}(\vec{\bf k})$ represents the eigenvector coefficient of an atomic orbital $\mu$. 
$\epsilon_j(\vec{\bf k})$ is the $j$th band energy at momentum $\vec{\bf k}$, 
$\vec{T}$ is the translational lattice vector, and $f_j(\vec{\bf k})$ is the occupation number of that state. 
In analogy to the density of states (DOS), for which the energy integration up to the Fermi energy gives the number 
of electrons, the energy integration of all COHP for a pair of atoms up to the Fermi energy (ICOHP) can imply 
the bond strength~\cite{M.Kupers2017}. 
All the COHP calculations are performed using the Local Orbital Basis Suite Towards Electronic-Structure 
Reconstruction (LOBSTER) code~\cite{V.L.Deringer2011,S.Maintz2013,S.Maintz2016} with the 
pbeVaspFit2015 basis set.\cite{S.Maintz2016}

The static exfoliation energy $E_{\rm exf}$ of a bulk MAB phase into 2D MBene is calculated, $e.g.$ for the case of 
M$_2$AlB$_2$,  
through $E_{\rm exf}= -\left[E_{\rm tot}({\rm M_2AlB_2})- 2 E_{\rm tot}({\rm M_2B_2})- 2 E_{\rm tot}({\rm Al})\right]/(4S)$, 
where $E_{\rm tot}({\rm M_2AlB_2})$, $E_{\rm tot}({\rm M_2B_2})$ and $E_{\rm tot}({\rm A})$ stand for 
the total energies of bulk M$_2$AlB$_2$ phase, 2D M$_2$B$_2$ MBene, and Al atom, respectively. 
The total energy of Al atom, $E_{\rm tot}({\rm Al})$, is 
estimated from its bulk structure. 
Here, $S$ is the surface area of the rectangular lattice, which is simply calculated by multiplication of $a$ and $c$ lattice parameters of the  M$_2$AlB$_2$ phase. 
Each unit cell of the M$_2$AlB$_2$ phase generates two MBene layers with totally 4 surfaces. 
Hence, the exfoliation energy is divided by 4 in the above formula.   
To calculate $E_{\rm exf}$ for other MAB phases, 
the above formula should be modified appropriately according to 
the number of Al atoms and the number of MBene layers in the cell.

 \begin{figure}[t]
\centering
\includegraphics[scale=0.42]{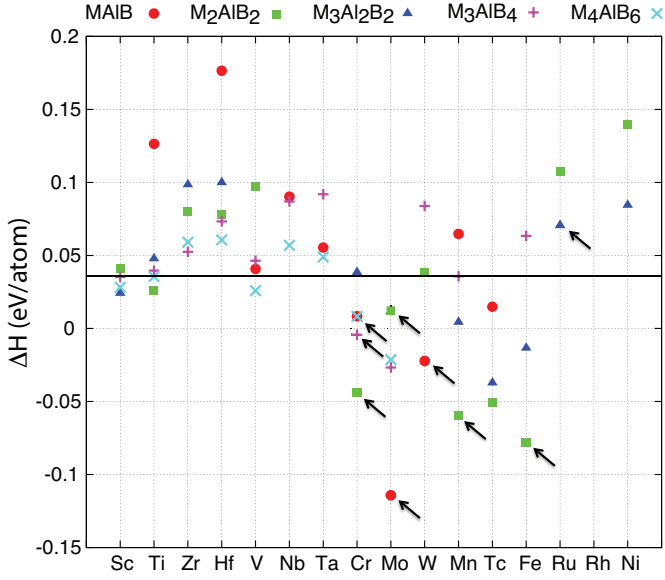}
  \caption{The relative formation energy $\Delta H$ of various MAB phases. The solid black line indicates 
  $\Delta H$ = 0.036 eV/atom. The arrows indicate the compounds that have already been synthesized experimentally.}
  \label{fig:enthalpy}
\end{figure}

\section{Results and discussion}

\subsection{Stabilities of MAB phases}

In order to find the trends in structural properties of MAB phases, we first investigate the dynamic stabilities of 
MAlB, M$_2$AlB$_2$, M$_3$Al$_2$B$_2$, M$_3$AlB$_4$, and M$_4$AlB$_6$ 
(M = Sc, Ti, Zr, Hf, V, Nb, Ta, Cr, Mo, W, Mn, Tc, Fe, Ru, Co, Rh, and Ni). 
Table~\ref{tab:data} lists the names and lattice parameters of MAB phases that we found dynamically stable 
based on the phonon calculations. There is no negative phonon frequency in the phonon spectra for these systems, 
shown in the supporting information file~\dag. All structures that have already been synthesized experimentally are 
among the MAB phases that we find stable in Table~\ref{tab:data}.  
The lattice parameters for the experimentally existing MAlB (M = Mo and W), M$_2$AlB$_2$ (M = Cr, Mn, and Fe), 
Cr$_3$AlB$_4$, and Cr$_4$AlB$_6$ are in excellent agreement with our theoretical results; the errors are less 
than 1.5\% except for the case of Ru$_3$Al$_2$B$_2$ in which the error is around 4.5\%. 
These results indicate the high quality of our calculations.

Next, in order to determine if the dynamically stable MAB phases found above 
can be realized experimentally, we calculate the formation energy $\Delta H$ of the MAB phases with 
respect to a combination of most competitive phases summarized in Table~\ref{tab:data}; 
$\Delta H = E_{\rm tot}({\rm MAB\ phase})-E_{\rm tot}({\rm competitive\ phases})$, where $E_{\rm tot}$ 
is the total energy. 
The most competing phases are found through the open quantum materials database (OQMD).\cite{R.Akbarzadeh2007,S.Kirklin2013} 
In order to observe the trend better, the $\Delta H$ is plotted with respect to transition metals in Figure~\ref{fig:enthalpy}.  
The negative (positive) $\Delta H$ indicates the relative stability (instability) of MAB phases with respect to the competitive phases. 
The more negative (positive) $\Delta H$ indicates the higher (smaller) chance for a synthesis of the MAB phases experimentally.
In excellent agreement with experimental observation, some of the Cr-, Mo-, W-, Mn-, and Fe-based MAB phases 
find negative $\Delta H$ (see Figure~\ref{fig:enthalpy}), indicating their synthesis possibility in 
experiment.\cite{M.Ade2015,P.Chai2015,S.Kota2016,T.N.Lamichhane2018} 
The negative $\Delta H$ of Tc-based MAB phases implies the high chance for their synthesis although there has been 
no experimental report.
This might be due to the lack of experiments on Tc-based MAB phases or there might be some binary Tc$-$Al or Tc$-$B, or 
ternary Tc$-$Al$-$B phases that have not been characterized experimentally yet, preventing the formation of Tc-based 
MAB phases. 
Further experiments on Tc$-$Al$-$B are highly desirable for the discovery of new Tc$-$Al, Tc$-$B, or Tc$-$Al$-$B phases.  

 \begin{figure}[t]
\centering
\includegraphics[scale=0.35]{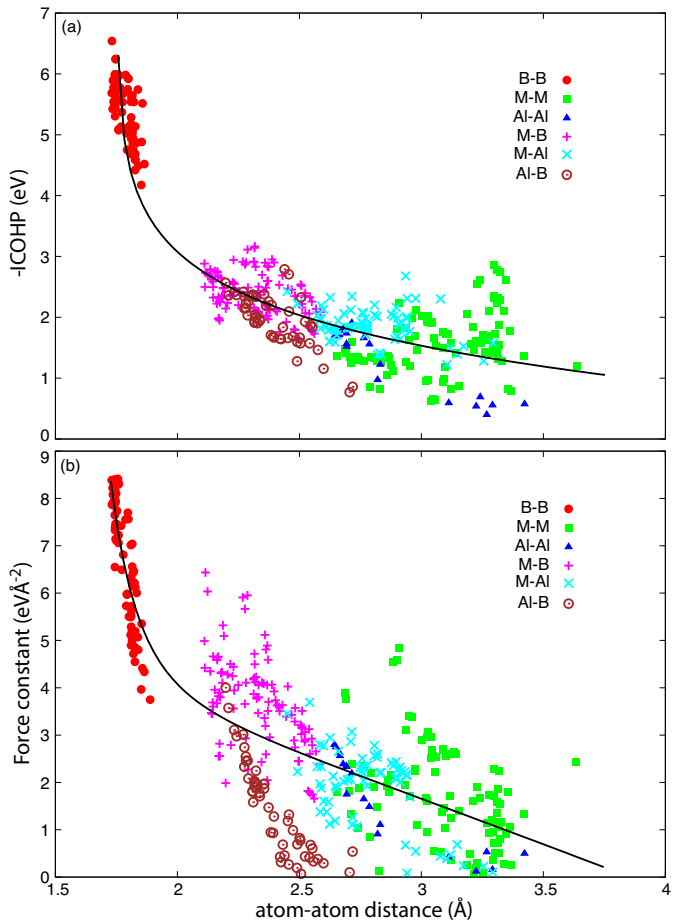}
  \caption{(a) integrated crystal orbital Hamilton population (ICOHP) and (b) force constants between various atoms 
  in MAB phases. 
  The solid lines are the guides to the eyes.
  }
  \label{fig:spring}
\end{figure}

Despite the experimental synthesis of Ru$_3$Al$_2$B$_2$,\cite{W.Jung1986} the calculated $\Delta H$ is a
relatively high positive value (0.07 eV/atom). 
This might imply that either our exchange-correlation function for Ru is not appropriate or
this compound could be formed experimentally because of the influences of vibrational entropy at high temperature and/or boron vacancies entropy effects~\cite{M.Dahlqvist2015}.   
According to the recent computational screening on inorganic crystal structure database (ICSD),\cite{G.Bergerhoff1983} 20\% of all compounds have instability of $\Delta H$ larger than 0.036 eV/atom.\cite{Y.Wu2013}
In such a case, Ru$_3$Al$_2$B$_2$ might be one of them. In other words, 80\% of experimentally synthesized 
compounds possess instability of $\Delta H$ less than 0.036 eV/atom.\cite{Y.Wu2013}
Assuming this value as a stability criterion ({\it i.e.}, $\Delta H$ $<$ 0.036 eV/atom), 
it is expected that there is a possibility for the experimental synthesis of Sc-, Ti-, and V-based MAB phases, 
in addition to Cr-, Mo-, W-, Mn-, Tc-, Fe-, and Ru-based MAB phases. In this study, we have focused on MAB phases with one type of transition 
metals. However, in the recent computational studies, it has been pointed out that it might be possible to synthesis ordered MAB phases
with double transition metals, M$_2$M$'$AlB$_4$ (M = Mn, Fe, Co, M$'$ = Cr, Mo, W).~\cite{F.Z.Dai2019}  

It is noteworthy that the synthesis of a pure MAB phase instead of multiple phases needs many trials. 
Experimentalists usually focus on transition metals that do not have many binary M$-$B or M$-$Al compounds. 
Then, one can try to improve the purity of a targeted phase by optimizing the experimental conditions such as temperature, 
the speed of quenching, and stoichiometry of mixture of transition metals, boron, and aluminum.


\subsection{Bond strength properties of MAB phases}

As described above, Al-containing layered materials are of great interest for obtaining novel 2D materials. 
In this regard, Al-containing MAB phases are a family of promising candidates for realizing 
2D transition metal borides. In MAB phases, MoAlB with double layers of Al has partially been etched into 2D MoB by removing Al atoms.\cite{L.T.Alameda2017,L.T.Alameda2018} 
Cr$_2$AlB$_2$ has also been etched completely into 2D CrB MBene.\cite{H.Zhang2018,H.Zhang2019} 
Current theoretical and experimental knowledge on exfoliation possibility of MAB phases is very limited. 

 The exfoliation is a dynamic chemical process that depends on the type of etchant, its concentration, 
the time duration of the treatment, and temperature. 
Similar to MAX phases, the obtained 2Ds from MAB phases will be covered with a mixture of O, OH, or F during exfoliation due to the chemical reactivity of transition metals.\cite{M.Khazaei2017, M.Khazaei2019} It has been shown that the surface termination can affect the stability of MXenes by forming chemical bonds between the chemical groups and the transition metals.\cite{M.Khazaei2017, M.Khazaei2019} In a similar way, it is expected the surface termination can affect the stability of MBene too. If one aims to simulate the exfoliation process using molecular dynamics simulations by including all above details, it will be very difficult and time-consuming and will almost be impossible to compare the exfoliation processes of different compounds or phases together. Therefore, the static calculations are almost the only efficient way to investigate or extract useful information about exfoliation possibility of layered materials. Here,
a set of static calculations are used to examine the bond strength properties
to systematically study the exfoliation possibility of MAB phases. 
This is because in a successful chemical process, the bond between an Al atom and its neighboring Al, M, and B  
atoms should be broken without destroying the other bonds. The weaker the M$-$Al bonds are, the higher 
the exfoliation possibility becomes. 
It is noteworthy that in our previous study,\cite{M.Khazaei2018_1} we successfully applied the same analysis on MAX phases. We predicted 37 new MAX phases that could be exfoliated. Later on, it was experimentally shown that Ti$_3$SiC$_2$, one of the phases that we also predicted, could be exfoliated into Ti$_3$C$_2$.\cite{M. Alhabeb2018}

  \begin{figure}[t]
\centering
\includegraphics[scale=0.39]{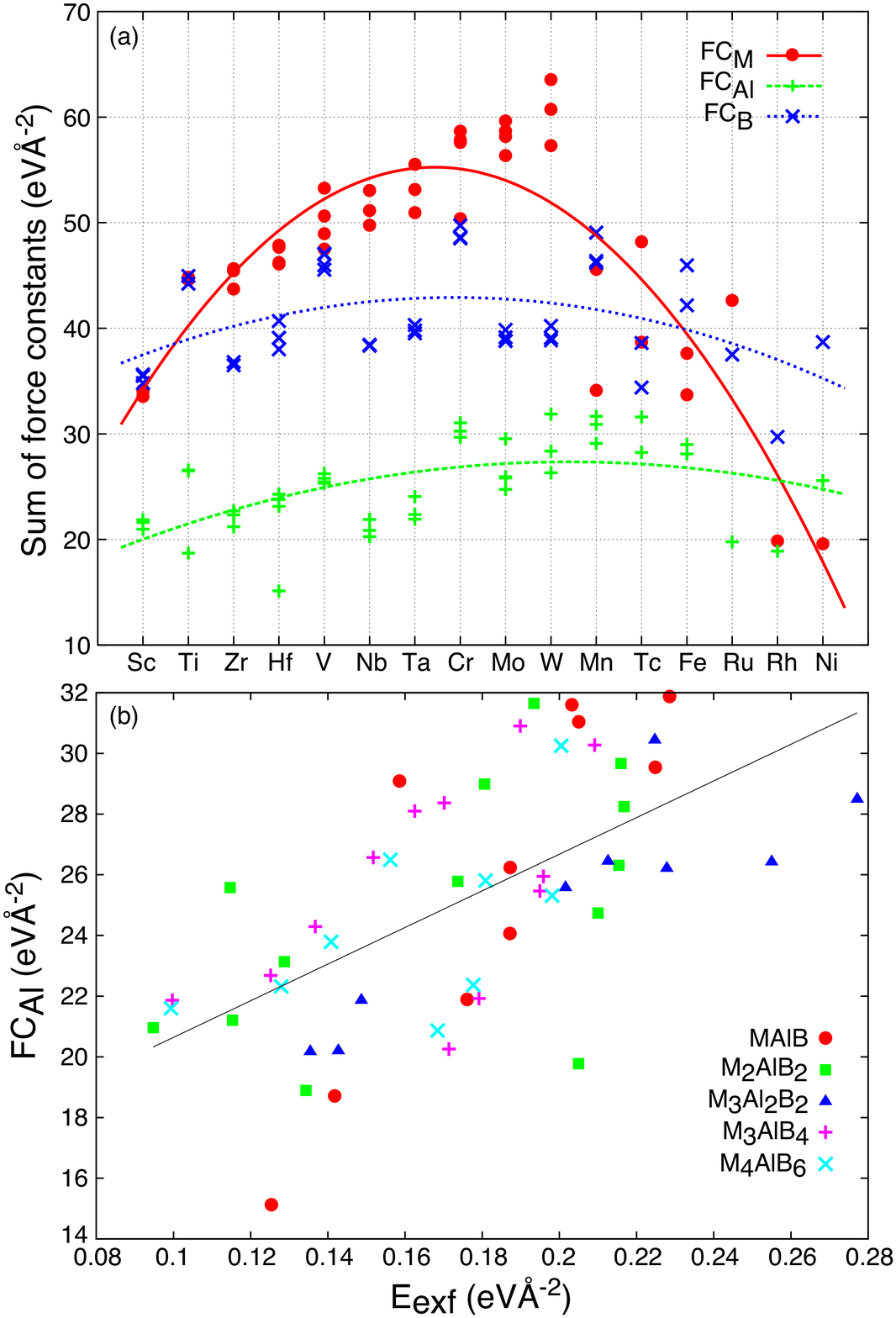}
  \caption{(a) The sum of the force constants for all bonds connected to an M, Al, and B atoms of various MAB phases. 
  The lines are the guides to the eyes.
  (b) FC$_{Al}$ versus the static exfoliation energy ($E_{\rm exf}$) for various MAB phases. 
  The correlation coefficient of the fitted line is r $=$ 0.622.
  }
  \label{fig:fc}
\end{figure}

The strength of a bond in a material can be approximately measured on the basis of 
the integrated COHP (ICOHP) analyses and the force constant calculations. 
The ICOHP has already been used to measure the bond strength in various materials.\cite{V.L.Deringer2011,S.Maintz2013,S.Maintz2016}
Previously, the force constants were used to measure the strength of bonds in diatomic dimers, molecules and clusters.\cite{R.G.Pearson1993,J.R.Lombardi2002,K.Brandhorst2008,D.Cremer2010} 
Recently, it has been expanded to study bond strengths in solids~\cite{M.Khazaei2018_1,Y.Liu2014,V.L.Deringer2015,J.Hong2016} and interfaces.\cite{O.V.Pupysheva2010,L.Wang2017}
We have also applied these two methods on MAX phase family. We have shown that generally, M$-$Al bonds in MAX phases 
are weaker than M$-$X bonds, which implies that M$-$Al bonds are relatively easily broken in MAX phases during 
the exfoliation process.\cite{M.Khazaei2018_1}   
However, it should be noted that neither of ICOHP nor force constant can perfectly describe the bond strength; 
the ICOHP mainly measures the strength of the covalency of a bond, but not its ionicity, while the force constant describes 
the behavior of energy potential when the bond is slightly stretched from its equilibrium size, but not its dissociation state.\cite{M.Khazaei2018_1}

Figure~\ref{fig:spring} summarizes the results of ICOHP and force constant calculations for various 
M$-$B, M$-$Al, M$-$M, Al$-$Al, Al$-$B, and B$-$B bonds in MAB phases.  
The results of both quantities show almost the same trends. Generally, as the atom$-$atom distance increases, 
the bond strength decreases. 
As shown in Fig.~\ref{fig:spring}, in general, the B$-$B and M$-$B bonds are stiffer than the M$-$Al, M$-$M, Al$-$Al, and 
Al$-$B bonds. This is consistent with the recent investigation on mechanical response of MoAlB and WAlB to 
external strain, finding that the cleavage occurs along the Al$-$Al layers due to the weak Al$-$Al  bonds.\cite{F.Z.Dai2018}  
It is noteworthy that exceptionally the Ni$-$Al is stiffer than Ni$-$B.
The B$-$B, M$-$B, and M$-$M bonds are the ones that construct the MB, M$_2$B$_3$, and M$_3$B$_4$ block units 
in MAB phases and are stronger than the M$-$Al, Al$-$Al, and Al$-$B bonds, implying the possibility of 
exfoliation of MAB phases by breaking the bonds between Al atoms and neighboring M, Al, and B atoms without 
hurting the 2D MB, M$_2$B$_3$, and M$_3$B$_4$ backbones, {\it i.e.,} MBenes. 
In MAX phases, the strong M$-$X (X = C or N) bonds construct the backbones, \textit{i.e.}, M$_{n+1}$X$_n$ MXenes. 
Our calculations indicate the B$-$B bonds in MAB phases are as strong as the M$-$X (X = C or N) bonds in MAX 
phases.\cite{M.Khazaei2018_1} This can be a sign that 2D MB, M$_2$B$_3$, and M$_3$B$_4$ may remain stable 
after the exfoliation process as in the case of M$_{n+1}$X$_n$ MXenes. 
The data in Figs.~\ref{fig:spring}(a) and (b) are shown for each phase separately in the supporting information file~\dag.

  \subsection{Exfoliation possibility of MAB phases into 2D MBenes}
  
 In order to better understand the exfoliation possibility of MAB phases, we also calculate the sum of the force 
 constants for all bonds connected to individual M, Al, or B atom (listed as FC$_M$, FC$_{Al}$, or FC$_B$ in 
Table~\ref{tab:data}) and the results are shown in Fig.~\ref{fig:fc}(a) 
 with respect to the transition metals in various MAB phases. 
 The FC$_M$ and FC$_ B$ are calculated for the M and B atoms that are closest to the Al atom. 
 The larger the FC$_M$ and FC$_ B$ are, the higher the chance is for obtaining perfect 2D MB, M$_2$B$_3$, 
 and M$_3$B$_4$ after the exfoliation process.
 The smaller FC$_{Al}$ indicates the ease for the breaking the bonds between an Al atom with its all neighboring atoms. 
 We have investigated the change of FC$_{Al}$ as a function of static exfoliation energy, shown in Fig.~\ref{fig:fc}(b). 
 As seen from this figure, the smaller (larger) FCAl appears to relate linearly with the smaller (larger) $E_{\rm exf}$ in each phase, 
although the linear correlation in M$_2$AlB$_2$ seems relatively lower than that in the other phases.
 A guideline of the overall trend between FC$_{Al}$ and $E_{\rm exf}$ is shown in Fig.~\ref{fig:fc}(b) with the linear correlation coefficient of 0.622. 
 This positive correlation can help us to explore exfoliation possibilities in an early study of the MAB materials.
 The data in Figs..~\ref{fig:fc}(a) and (b) are shown for each phase separately in the supporting information file~\dag.

 \begin{figure}[t]
\centering
\includegraphics[scale=0.32]{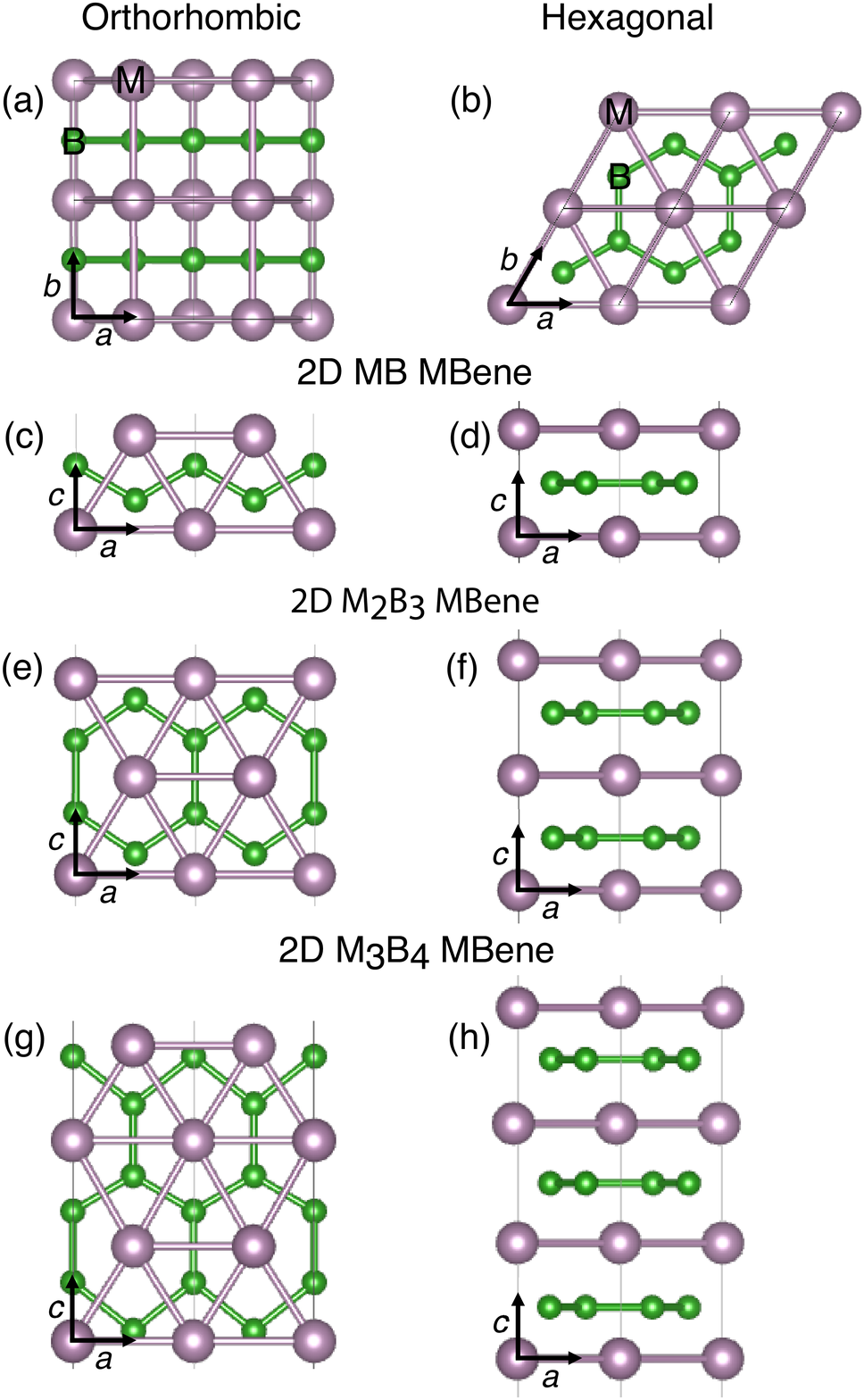}
  \caption{Top views of (a) rectangular and (b) hexagonal 2D MB, M$_2$B$_3$, and M$_3$B$_4$ MBenes. 
  (c), (e), and (g) [(d), (f), and (h)] Side views of rectangular (hexagonal) 2D 
  MB, M$_2$B$_3$, and M$_3$B$_4$, respectively. 
  $a$, $b$, and $c$ directions are indicated by arrows.
  The structures are periodic in $ab$ plane and vacuum space of 30~\AA~exists along the $c$ axis. 
  }
  \label{fig:mbene}
\end{figure}

 As shown in Fig.~\ref{fig:fc}(a), in most of MAB phases, FC$_M$ $>$ FC$_ B$ $>$ FC$_{Al}$. 
 We also find that FC$_M$, FC$_ B$, and FC$_{Al}$ continuously increase when the type of transition metal changes 
 from Sc$\rightarrow$W, while FC$_M$ and FC$_ B$ start decreasing when the transition metal is changed from 
 W$\rightarrow$Ni.  
 The family of Cr-, Mo-, and W-based MAB phases obtain almost the highest FC$_M$ and FC$_ B$, 
 implying that these MAB phases may have large elastic constants and there will be a good chance for obtaining 
 perfect 2D MB, M$_2$B$_3$, and M$_3$B$_4$ MBenes. In Mn$\rightarrow$Ni MAB phases, 
 the FC$_M$, FC$_ B$, and FC$_{Al}$ start decreasing. 
Therefore, there is a less chance to obtain perfect MBenes in these MAB phases because 
 the etchant can wash out not only the Al atoms but also M and B atoms. 
Our calculations predict that when the transition metal changes from Sc$\rightarrow$W (Mn$\rightarrow$Ni), the 
 degree of mechanical anisotropy of MAB phases increases (decreases).

Experimentally, MoAlB with double layers of Al atoms has been partially etched into 2D MoB MBene\cite{L.T.Alameda2017,L.T.Alameda2018} 
and Cr$_2$AlB$_2$ has also been etched completely into 2D CrB MBene.\cite{H.Zhang2018,H.Zhang2019}
In the family of Cr-, Mo-, and W-based MAB phases, we find in Fig.~\ref{fig:fc}(a) that the bonding between Al atoms 
and other atoms are very strong. Indeed, the total force constant FC$_{Al}$ for MoAlB is 29.542 eV/\AA$^2$ 
(see Table~\ref{tab:data}). 
This implies that if experimentalists were able to exfoliate MoAlB partially, the full or partial exfoliation of all MAB 
phases from Sc$\rightarrow$W might also be possible. This is consistent with the experimental results that experimentalists succeeded  
to exfoliate Cr$_2$AlB$_2$ into 2D CrB nanosheets.\cite{H.Zhang2018,H.Zhang2019}

Unfortunately, the experimental data on the exfoliation of 
MAB phases is still very limited. However, there are many on the exfoliation of MAX phases. 
By investigating MAX phases, we have found previously that MAX phases already exfoliated into 2D MXenes 
satisfy FC$_M$ $>$ 44.253 , FC$_ {Al}$ $<$ 21.855, FC$_ X$ (X = C or N) $>$ 40.511 eV/\AA$^2$.\cite{M.Khazaei2018_1} 
By comparing the FC$_ {Al}$ values of MAB phases with those of the exfoliated MAX phases, 
we find that the FC$_{ Al}$ values in MAB phases are larger than those in MAX phases. 
This suggests that the exfoliation of MAB phases is more difficult than MAX phases and explains why MoAlB 
was not completely exfoliated experimentally. 
As a strategy to enhance the exfoliation possibility of MoAlB, it might be appropriate to alloy MoAlB with other transition 
metals, $e.g.$, M$'$ = Sc, Ti, Zr, Hf, V, Nb, or Ta in which their corresponding MAB phases possess lower
FC$_{Al}$ values. In this way, removing  Al atoms from Mo$_x$M$'_{1-x}$AlB phase might become easier.

Our force constant data in Fig.~\ref{fig:fc}(a) indicate that depending on the transition metal, 
some of the phases possess the FC$_{Al}$ of almost in the same range, therefore all 
those phases may have similar chances for Al removal. 
However, as explained before, the perfectness of the obtained 2D MBenes will depend on FC$_M$ and FC$_B$.  

\subsection{Structural phase transformation of 2D MBenes}   

In 2D M$_{n+1}$X$_n$ MXenes, it has been observed experimentally that upon heating, the 2D MXenes transform 
into nanocrystals of transition metal oxides and disordered graphitic carbon structures.\cite{M.Naguib2014} 
It is also known that some of the transition metals, M = Sc, Ti, V, Nb, Ta, Cr, Mo, and W, have MB$_2$ hexagonal 
structures.\cite{G.Akopv2017} Therefore, it might be plausible that 
orthorhombic 2D MB, M$_3$B$_4$, and M$_2$B$_3$ MBenes with rectangular lattice structures transform into monolayer or 
multilayers of graphene-like boron sheets sandwiched between transition metals, as shown in Fig.~\ref{fig:mbene}.
The orthorhombic 2D MB, M$_3$B$_4$, and M$_2$B$_3$ contain zigzag single, double, or triple chains of boron atoms, respectively. In M$_3$B$_4$ and M$_2$B$_3$, 
the zigzag chains form strips of boron hexagonal networks. During the transformation from rectangular lattice to a hexagonal lattice, the zigzag chains start bonding to each other on $ab$ plane, and consequently the chains or hexagonal strips transform into single, double, and triple graphene-like boron sheets shown in Fig.~\ref{fig:mbene}. In order to determine which of the 2D MBenes with orthorhombic or hexagonal lattices are preferable for a particular application, additional calculations will be necessary.

\begin{figure}[t]
\centering
\includegraphics[scale=0.35]{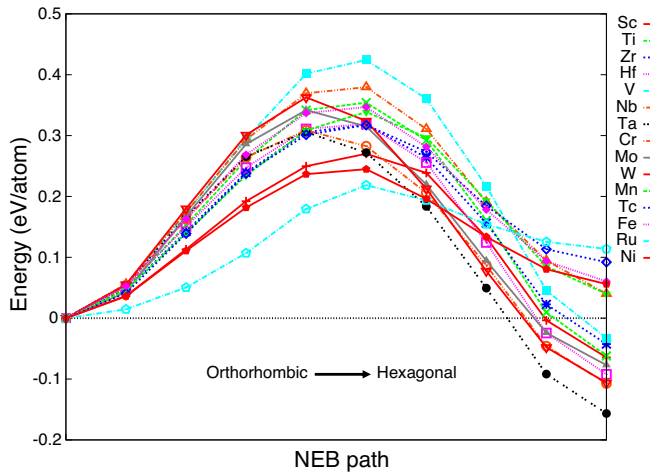}
  \caption{The energy barrier for structural transformation of 2D MB MBenes from the rectangular lattice to 
  hexagonal lattice structures calculated using the nudged elastic band (NEB) method.   
  }
  \label{fig:neb}
\end{figure}

Here, we calculate the energy barrier for such a transformation for the 2D MB MBenes from the orthorhombic to 
a hexagonal phase using nudged elastic band (NEB) method,\cite{G.Henkelman2000_1,G.Henkelman2000_2}. 
The results are shown in Fig.~\ref{fig:neb}. 
We find that the 2D MB (M = Sc, Ti, Zr, Hf, V, Nb, Ta, Mo, and W) MBenes with hexagonal lattice structures are 
more stable than the orthorhombic phases. In contrast, we find that the orthorhombic phases are more stable for 
M = Cr, Mn, Tc, Fe, Ru, and Ni. 
This is reasonable because Sc, Ti, Zr, Hf, V, Nb, Ta, Mo, and W are found experimentally to form in hexagonal binary 
boride phases.\cite{G.Akopv2017} Our NEB calculations indicate that the energy barrier for transformation 
from the orthorhombic to hexagonal phase is between 0.2$-$0.4 eV/atom. However, it should be noted that 
this energy barrier is overestimated because it is very difficult to consider the possibility of nucleation of the boron 
sheets in our calculations.    

\section{Conclusion}

 Recently, a layered family of Al-containing transition metal borides, known as MAB phases has received attention for 
 obtaining 2D transition metal borides, {\it i.e.,} 2D MB, M$_2$B$_3$, and M$_3$B$_4$ MBenes. 
 Currently, there exist only a few members of MAB phases, mainly made of Cr, Mo, W, Mn, and Fe.
 Here, based on the phonon and formation energy calculations, we have shown that in addition to these MAB phases, 
 Sc, Ti, V, and Tc-based MAB phases are also possible to be realized. The bond strength analyses indicate the 
 B$-$B and M$-$B bonds are stronger than M$-$Al, Al$-B$, and Al$-$Al bonds, implying the successful exfoliation 
 possibility of MAB phases into 2D MBenes. Our analyses indicate that 
  Cr-, Mo-, and W-based MAB phases are the best candidates for obtaining 2D MBenes. 
  To enhance the exfoliation possibility of Cr-, Mo-, and W-based MAB phases,
  we suggest to alloy these MAB phases with Sc, Ti, Zr, Hf, V, Nb, or Ta.  
 Furthermore, we have investigated 
 the structural transformation of 2D MBenes from the orthorhombic to hexagonal phase, 
 which results in the formation of graphene-like boron sheets sandwiched between transition metals. 
 This transformation might be possible at high temperatures especially for the case of Sc-, Ti, Zr-, Hf-, V-, Nb-, Ta-, Mo-, and W-based MBenes.

\section*{Conflicts of interest}
There are no conflicts to declare.

\section*{Acknowledgements}
M.K. is grateful to RIKEN Advanced 
Center for Computing and Communication (ACCC) for the allocation of computational resource of the RIKEN 
supercomputer system (HOKUSAI GreatWave). Part of the calculations were also performed on Numerical 
Materials Simulator at National Institute for Materials Science (NIMS). M.K. gratefully acknowledges the support by Grant-in-Aid for Scientific Research 
(No. 17K14804) from MEXT Japan.





\end{document}